\documentstyle[aps,multicol,epsfig]{revtex}

\begin{document}
\draft
\preprint{}
\title{Reexamination of $d$-wave superconductivity in the
two-dimensional Hubbard model
}
\author{Takao Morinari}
\address{Yukawa Institute for Theoretical Physics, Kyoto University
Kyoto 606-8502, Japan
}
\date{\today}
\maketitle
\begin{abstract}
We reexamine the possibility of $d$-wave superconductivity in the
hole-doped two-dimensional Hubbard model.
In terms of the gauge field description of the spin fluctuations,
we show that $d$-wave superconductivity is unstable in the
perturbative region with respect to the on-site Coulomb repulsion,
$U$. Whereas in the region where d-wave superconductivity is possible,
there is a strong constraint on the gap of superconductivity.
Analysis of the localized spin moments suggests that there is another
$d$-wave ($d_{x^2-y^2}$-wave) pairing due to the short-range
antiferromagnetic correlation.
\end{abstract}

\begin{multicols}{2}

\narrowtext

Over the past few decades a considerable number of studies have been
made on the Hubbard model.
In spite of its simplicity, we can expect a variety of phenomena:
antiferromagnetism, ferromagnetism, and unconventional
superconductivity.
Among others, the possibility of $d$-wave superconductivity has been
considerably investigated since the discovery of the high $T_c$ 
cuprates.\cite{SCALAPINO}

The theory of $d$-wave superconductivity in the Hubbard model is based
on the antiferromagnetic spin fluctuations.\cite{SCALAPINO}
The mechanism is analogous to the p-wave pairing
mechanism due to the ferromagnetic spin fluctuations in the $^3$He
system.\cite{He3}
In principle, the theory relys on the perturbative expansion with
respect to $U$, the on-site Coulomb repulsion.

However, the theory has some disadvantages.
First of all, it is hard to describe the half-filled case in which
local spin moments are produced at each site and the interaction
between the spin moments is proportional to the inverse of $U$.
Apparently, this dependence of the interaction on $U$ cannot be
reproduced by the perturbative expansion in $U$.
In addition, realistic values of $U$ are much larger than the
carrier hopping amplitude, $t$. 
Therefore, we must pay particular attention in applying the
perturbative expansion in $U$, especially when we deal with the effect
of the localized spin moments.

In this paper, we re-examine the possibility of $d$-wave
superconductivity in the hole-doped two-dimensional Hubbard model
without relying on the perturbative expansion with respect to $U$.
Starting from the path-integral form of the partition function, we
introduce the Stratonovich-Hubbard field for the value of the
localized spin moments.
Analysis of it suggests that superconductivity based on the
antiferromagnetic spin fluctuations is unstable in the perturbative
region.
While in the region where $d$-wave superconductivity is possible, 
we show that there is a strong
constraint on the value of the superconducting gap.
Instead of the $d$-wave Cooper pairing, there is another
$d$-wave ($d_{x^2-y^2}$-wave) pairing due to the antiferromagnetic
short-range correlation.
This $d_{x^2-y^2}$-wave pairing state is similar to the spinon
pairing in the RVB theory.\cite{ANDERSON_ETAL,KOTLIAR_LIU}
However, the crucial difference here is that we do not need the $U(1)$ 
gauge symmetry breaking that is essential for the slave-particle gauge 
theory.\cite{NAYAK}

{\it Formulation-}
The partition function of the Hubbard model is written as
${\cal Z}=\int {\cal D}\overline{c} {\cal D} c \exp (-S)$, where
\begin{eqnarray}
S&=&\int_0^{\beta} d\tau \left[ 
\sum_j \overline{c}_j \left( \partial_{\tau} - \mu \right) c_j
-t\sum_{\langle i,j \rangle} \left( \overline{c}_i c_j + 
\overline{c}_j c_i
\right) 
\right. \nonumber \\ & & \left. 
+ U \sum_j n_{j\uparrow} n_{j\downarrow} \right].
\end{eqnarray}
Here $\tau$ dependence of all fields is implicit and
the summation $\sum_{\langle i,j \rangle}$ is taken over the
nearest neighbor sites.
Carrier fields are represented in  spinor representation:
$c_i = ^T \left( \begin{array}{cc}
c_{i\uparrow} & c_{i\downarrow} \end{array} \right)$ 
and 
$\overline{c}_i = \left( \begin{array}{cc}
\overline{c}_{i\uparrow} & \overline{c}_{i\downarrow}\end{array}
\right)$.
The on-site Coulomb interaction term can be rewritten as,
$U\sum_j n_{j\uparrow} n_{j\downarrow} = (U/4) \sum_j
\left[ (n_{j\uparrow} + n_{j\downarrow} )^2
- \left( \overline{c}_j 
{\mbox{\boldmath ${\bf \sigma}$}} c_j \right)^2 
+ 2 ( n_{j\uparrow} + n_{j\downarrow} )\right]$,
where the components of the vector 
${\mbox{\boldmath ${\bf \sigma}$}} = (\sigma_x,\sigma_y,\sigma_z)$ 
are the Pauli spin matrices.
Introducing Stratonovich-Hubbard fields for the charge and spin
degrees of freedom,\cite{SCHULZ} we obtain $Z=\int {\cal D}\overline{c}
{\cal D} c {\cal D} {\mbox{\boldmath ${\bf \Omega}$}} {\cal D}\phi_{c} 
{\cal D}\phi \exp (-S)$, where the action is given by
\begin{eqnarray}
S&=&
\int_0^{\beta} d\tau \left[
\sum_j \overline{c}_j \left( \partial_{\tau} -\mu \right) c_j
-t \sum_{\langle i,j \rangle} \left( \overline{c}_i c_j +
\overline{c}_j c_i \right) 
\right. \nonumber \\ & & \left. 
+ \frac{U}{4} \sum_j \phi_j^2 
- \frac{U}{2} \sum_j \phi_j 
{\mbox{\boldmath ${\bf \Omega}$}}_j \cdot
\overline{c}_j {\mbox{\boldmath ${\bf \sigma}$}} c_j
\right. \nonumber \\ & & \left. 
+ \frac{U}{4} \sum_j \phi_{cj}^2 
- \frac{iU}{2} \sum_j \phi_{cj} \overline{c}_j c_j  \right],
\end{eqnarray}
up to constant.
Here the vector ${\mbox{\boldmath ${\bf \Omega}$}}_j$ is a unit
vector,  $\phi_j$ represents the value of the localized
spin moments, and  $\sigma_0$ is the unit matrix in spin space .
The scalar $\phi_{cj}$ is associated with the charge fluctuations.
Note that $\phi_j = 1$ at half-filling. If we consider hole doping,
then $\phi_j$ takes $\phi_j \leq 1$.
For these amplitudes, we do not consider the possibility of
inhomogeneous configurations of them because such configurations may
compete with superconductivity.
In order to focus on the possibility of superconductivity, we
take the uniform values, that is,
$\phi_j = \phi = {\rm const.}$ and
$i\phi_{cj}= - \langle \overline{c}_j \sigma_0 c_j \rangle = -
(1-\delta)$, with $\delta$ the doped hole concentration.
As a result, we may write the action in the following form
\begin{eqnarray}
S&=&\int_0^{\beta} d\tau \left[
\sum_j \overline{c}_j \left( \partial_{\tau} -\tilde{\mu} \right) c_j
-t \sum_{\langle i,j \rangle} \left( \overline{c}_i c_j +
\overline{c}_j c_i \right) 
\right. \nonumber \\ & & \left. 
- \frac{\phi U}{2} \sum_j 
{\mbox{\boldmath ${\bf \Omega}$}}_j \cdot
\overline{c}_j {\mbox{\boldmath ${\bf \sigma}$}} c_j
\right]
+ \frac{\phi^2 U}{4} \beta N,
\label{eq_S}
\end{eqnarray}
where $\tilde{\mu}=\mu-U(1-\delta)/2$ and $N$ is the number of the
lattice sites.
Note that the last term in the square brackets has the form of
Hund coupling between the localized spin moments and the carriers' spins.

{\it Gauge field description of the spin fluctuations-}
Now we rotate the spin of the fermion at $j$-site so that it is in the 
direction of ${\mbox{\boldmath ${\bf \Omega}$}}_j$.
Such rotation can be done by performing the following SU(2)
transformation:
\begin{equation}
\left\{
\begin{array}{c}
c_j = U_j f_j, \\
\overline{c}_j = \overline{f}_j \overline{U}_j,
\end{array}
\right.
\label{eq_SU2}
\end{equation}
where,
$U_j = \left(
\begin{array}{cc}
z_{j\uparrow} & -\overline{z}_{j\downarrow} \\
z_{j\downarrow} & \overline{z}_{j\uparrow} 
\end{array}
\right)$
and 
$~\overline{U}_j = \left(
\begin{array}{cc}
\overline{z}_{j\uparrow} & \overline{z}_{j\downarrow} \\
-z_{j\downarrow} & z_{j\uparrow} 
\end{array}
\right).
$
Here complex variables $\overline{z}_{j\sigma}$ and $z_{j\sigma}$ are
defined as
${\mbox{\boldmath ${\bf \Omega}$}}_j =
\overline{z}_j 
{\mbox{\boldmath ${\bf \sigma}$}} z_j$.\cite{SCHULZ}
Thus, we obtain
\begin{eqnarray}
S&=& \int d\tau \left[ \sum_j \overline{f}_j \left( \partial_{\tau} -
\tilde{\mu} + \overline{U}_j \partial_{\tau} U_j \right) f_j
\right. \nonumber \\ & & \left. 
-t \sum_{\langle i,j \rangle} \left( 
\overline{f}_i \overline{U}_i U_j f_j + 
\overline{f}_j \overline{U}_j U_i f_i \right) 
\right. \nonumber \\ & & \left. 
- \frac{\phi U}{2} \sum_j
\overline{f}_j \sigma_z f_j \right] + \frac{\phi^2 U}{4} \beta N.
\label{eq_Sf}
\end{eqnarray}
The effective action of the boson fields $\overline{z}_j$ and $z_j$ is 
obtained by integrating out fermion fields $\overline{f}_j$ and $f_j$:
\begin{eqnarray}
S_{\rm eff} &=& - {\rm Tr} \ln \left[ 
\left( \partial_{\tau} - \tilde{\mu} - \frac{\phi U}{2} \sigma_z
\right)
\delta_{ij} 
\right. \nonumber \\ & & \left. 
- t_{ij} \overline{U}_i U_j + \overline{U}_i \partial_{\tau} U_i
\delta_{ij}
\right] + \frac{\phi^2 U}{4} \beta N,
\label{eq_Sz}
\end{eqnarray}
where $t_{ij}=t$ for the nearest neighbor sites and $t_{ij}=0$
otherwise.
If we expand Eq.~(\ref{eq_Sz}) with respect to $\phi U$, then the
second order term yields the RKKY interaction between the localized
spin moments.\cite{LACOUR-GAYET_CYROT}
On the other hand, the second order of the expansion of
Eq.~(\ref{eq_Sz}) with respect to $t_{ij}$ yields the
antiferromagnetic Heisenberg
Hamiltonian.\cite{SCHULZ,LACOUR-GAYET_CYROT}

Before going into the analysis of the action (\ref{eq_Sf}), we
discuss how the formulation is related to the antiferromagnetic spin
fluctuation theory that is based on the perturbative expansion in
$U$\cite{SCALAPINO}.
Formally, we can reproduce the effective interaction of the
antiferromagnetic spin fluctuation theory as follows.
Expanding the action (\ref{eq_Sz}) in powers of $U$, the quadratic
term yields the boson propagator.
The effective interaction due to the exchange of the bosons
corresponds to the interaction derived from the antiferromagnetic
spin-fluctuation theory \cite{SCALAPINO} up to the paramagnon
contributions that leads to $p$-wave pairing.\cite{He3}
Apparently, we can trust this interaction only in the perturbative
region with respect to $U$.
In addition, the interaction exists only when $\phi \neq 0$, or
coupling to the bosons is lost.
Nevertheless we will show later that $\phi=0$ in the perturbative
region in $U$.

For the analysis of the system,
we rely on neither the $\phi U$-expansion nor the $t$-expansion
because both of them are reliable only in part of the parameter range
of $U/t$ and $\delta$.
Alternatively, we study the system by taking the continuum limit.
Taking such limit is justified when the fluctuations are
long-ranged. Since the antiferromagnetic spin fluctuations may be
long-ranged near half-filling, we may take the continuum limit.
In the continuum limit, the action (\ref{eq_Sf}) is reduced to 
\begin{eqnarray}
S &=& \int d \tau \int d^2 {\bf r}
\overline{\psi} ({\bf r},\tau) 
\left[ 
\left( \partial_{\tau} - \tilde{\mu} \right) \sigma_0
+i {\cal A}_{\tau} - \frac{\phi U}{2}\sigma_z
\right. \nonumber \\ & & \left. 
- \frac{1}{2m}
\left( -i \sigma_0 \nabla + 
{\mbox{\boldmath ${\bf {\cal A}}$}} \right)^2
\right]
\psi ({\bf r},\tau)
+ S_{\cal A},
\label{eq_Sa}
\end{eqnarray}
where the SU(2) gauge field ${\cal A}_{\mu}$ is defined as
${\cal A}_{\mu} = \sum_{a=x,y,z} {\cal A}^a_{\mu} \sigma_a = 
-i \overline{U} \partial_{\mu} U$ and
$S_{\cal A}$ is derived from Eq.~(\ref{eq_Sz}) in principle.

The system governed by the action (\ref{eq_Sa}) is the fermion system
with the interaction due to the exchange of the SU(2) gauge field 
${\cal A}_{\mu}$.
Here the gauge field ${\cal A}_{\mu}$ is associated with the spin
fluctuations.
The $z$-component ${\cal A}^z_{\mu} = -i (\overline{z}_{\uparrow} 
\partial_{\mu} z_{\uparrow}
+ \overline{z}_{\downarrow} \partial_{\mu} z_{\downarrow} )$ 
describes the ferromagnetic spin
fluctuations. 
Whereas the $x$-component ${\cal A}^x_{\mu} = 
-i (z_{\uparrow} \partial_{\mu} z_{\downarrow} - z_{\downarrow}
\partial_{\mu} z_{\uparrow} )$ describes the antiferromagnetic spin
fluctuations.
The latter relationship is implied from the analysis of the $CP^1$
representation of the antiferromagnetic Heisenberg
model.\cite{READ_SACHDEV}

In order to consider the antiferromagnetic spin fluctuations,
we focus on the gauge field ${\cal A}_{\mu}^x$.
In terms of the fields $\tilde{\psi}_{\pm} = (\psi_{\uparrow} \pm
\psi_{\downarrow})/\sqrt{2}$ that diagonalize the gauge charge,
the action is rewritten as
\begin{eqnarray}
S &=& \int d \tau \int d^2 {\bf r} \left\{
\sum_{s=\pm} \overline{\tilde{\psi}}_s ({\bf r},\tau) 
\left[
\partial_{\tau} - \tilde{\mu}
+i s {\cal A}^x_{\tau} 
\right. \right. \nonumber \\ & & \left. \left. 
- \frac{1}{2m} \left( -i \nabla + 
s {\mbox{\boldmath ${\bf {\cal A}}$}}^x \right)^2 
\right] \tilde{\psi}_s ({\bf r},\tau) \right. \nonumber \\
& & \left. - \frac{\phi U}{2} 
\left[
\overline{\tilde{\psi}}_+ ({\bf r},\tau)
\tilde{\psi}_- ({\bf r},\tau)
+ 
\overline{\tilde{\psi}}_- ({\bf r},\tau)
\tilde{\psi}_+ ({\bf r},\tau)
\right] \right\}
+ S_{{\cal A}^x}.
\label{eq_Sa2}
\end{eqnarray}
The action (\ref{eq_Sa2}) has the form of fermions coupled with
the U(1) gauge field.

{\it Analysis of the gap equation-}
Now we study the possibility of spin singlet superconductivity based
on the action (\ref{eq_Sa2}).
In the following we assume that there exists an attractive interaction 
induced by the exchange of the gauge field.
In the presence of an attractive interaction, the mean field
Hamiltonian for the spin singlet pairing state may have the following
form
\begin{eqnarray}
H_{\rm MF}&=& \frac12
\sum_{\bf k} 
\left(
\begin{array}{cccc}
\tilde{c}_{{\bf k},+}^{\dagger} &
\tilde{c}_{{\bf k},-}^{\dagger} &
\tilde{c}_{-{\bf k},+} &
\tilde{c}_{-{\bf k},-}
\end{array}
\right) 
\nonumber \\ & & \times
\left( 
\begin{array}{cccc}
\xi_k & -\phi U/2 & 0 & \Delta_{\bf k} \\
-\phi U/2 & \xi_k & - \Delta_{\bf k} & 0 \\
0 & - \Delta_{\bf k}^* & -\xi_k & \phi U/2 \\
\Delta_{\bf k}^* & 0 & \phi U/2 & - \xi_k 
\end{array}
\right)
\left(
\begin{array}{c}
\tilde{c}_{{\bf k},+} \\
\tilde{c}_{{\bf k},-} \\
\tilde{c}_{-{\bf k},+}^{\dagger} \\
\tilde{c}_{-{\bf k},-}^{\dagger}
\end{array}
\right).
\end{eqnarray}
The gap $\Delta_{\bf k}$ is evaluated from the gap equation:
$\Delta_{\bf k} 
=
-\frac{1}{4\Omega} \sum_{\bf k}
V_{{\bf k}{\bf k}'} \frac{\Delta_{{\bf k}'}}{E_{{\bf k}'}}
\left[ \tanh \frac{\beta (E_{{\bf k}'}+\phi U/2)}{2} 
+ \tanh \frac{\beta (E_{{\bf k}'}-\phi U/2)}{2}
\right]$,
where $E_{\bf k}=\sqrt{\xi_k^2+|\Delta_{\bf k}|^2}$.
In principle, the interaction $V_{{\bf k}{\bf k}'}$ is derived from
Eqs.~(\ref{eq_Sz}) and (\ref{eq_Sa})
by eliminating the gauge field ${\cal A}_{\mu}^x$.
However, we do not need its explicit form.

At zero temperature, the gap equation is reduced to 
\begin{equation}
\Delta_{\bf k} = -\frac{1}{2\Omega} \sum_{{\bf k}',E_{{\bf k}'}>\phi
U/2}
V_{{\bf k}{\bf k}'} \frac{\Delta_{{\bf k}'}}{E_{{\bf k}'}},
\label{eq_gap}
\end{equation}
Note that in Eq.~(\ref{eq_gap}) the summation in ${\bf k}'$-space is
taken over under the constraint $E_{{\bf k}'} > \phi U/2$.
The presence of the constraint on spin singlet pairing states is
understood as follows.
One can see that the second term in the braces in Eq.~(\ref{eq_Sa2}) 
is similar to that of the Zeeman energy term produced by applying an
in-plane magnetic field to the system.
In fact, such Zeeman energy term is proportional to the applied
magnetic field times $\sum_j (c_{j\uparrow}^{\dagger} c_{j\downarrow}
+ c_{j\downarrow}^{\dagger} c_{j\uparrow})$. (Here the direction of
the in-plane magnetic field is chosen along the $x$-axis.)
Apparently, in the limit of the large in-plane magnetic field, spin
singlet pairing sates are unstable. Similarly, in the large $\phi U$
limit, spin singlet states are not stable.
Therefore, if spin singlet superconductivity is stable, then the
superconducting gap $\Delta$ should satisfy,
\begin{equation}
\Delta > \phi U/2 \equiv \Delta_c.
\label{eq_cond}
\end{equation}
In order to find the doped hole concentration dependence of the
constraint, we need to evaluate $\phi$.
For the calculation, 
we assume the staggered form for 
${\mbox{\boldmath ${\bf \Omega}$}}_j$ as
${\mbox{\boldmath ${\bf \Omega}$}}_j = (-1)^j {\hat e}_z$,
because we are concerned with the antiferromagnetic spin
fluctuations.\cite{SPIRAL}
Note that non-zero value of $\phi$ does not imply the presence of the
antiferromagnetic long-range order but it just implys the presence of
the antiferromagnetic correlation because there is the phase
fluctuations, or the effect of the gauge field ${\cal A}^x_{\mu}$, 
as well as the quantum fluctuations.

Now we estimate the value of $\phi$ by solving the saddle point
equations derived from the action obtained from Eq.~(\ref{eq_S}) after
integrating out $\overline{c}_j$ and $c_j$.
The variation with respect to $\phi$ yields
\begin{equation}
\frac{U}{4\pi^2 t} \int^1_{\sqrt{\nu^2-\alpha^2}}
d\gamma \frac{K(\sqrt{1-\gamma^2})}{\sqrt{\gamma^2+\alpha^2}} =1,
\label{eq_al}
\end{equation}
where $\alpha=\phi U/(8t)$, $\nu=\tilde{\mu}/(4t)$, and 
$K(\xi)=\int_0^{\pi/2} d\theta (1/\sqrt{1-\xi^2 \sin^2 \theta})$ 
is the complete elliptic integral of the first kind. 
Meanwhile the variation with respect to $\mu$ yields
\begin{equation}
\frac{4}{\pi^2} \int^1_{\sqrt{\nu^2-\alpha^2}} d\gamma
K(\sqrt{1-\gamma^2}) = 1 - \delta.
\label{eq_del}
\end{equation}
In deriving these equations, we have used $\nu <0$ and $|\nu|>\alpha$,
which holds for the hole doped case.

\begin{figure}[tbhp]
\center
\epsfxsize=2.0truein
\psfig{file=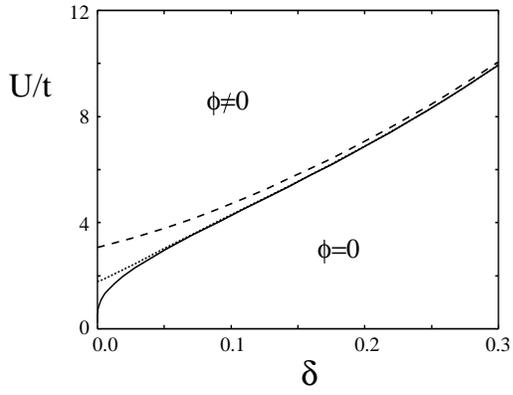,width=2.0in,angle=270}
\vspace{0.1in}
\caption{The boundary between $\phi \neq 0$ and $\phi=0$
in the $U/t-\delta$ plane. The boundary is given by the solid curve.
The dashed curve represents $\Delta_c/t=0.20$ and the dotted
curve $\Delta_c/t=0.04$.
}
\label{fig_af}
\end{figure}

From Eqs.~(\ref{eq_al}) and (\ref{eq_del}) we find $U/t$ and $\delta$
dependence of $\phi$ and $\Delta_c$.
The boundary between $\phi \neq 0$ and $\phi=0$ is shown in
Fig.~\ref{fig_af} by the solid curve.
In the $\phi = 0$ regime, there is no attractive interaction due to
the absence of the antiferromagnetic spin fluctuations as mentioned
above.
Note that the $\phi = 0$ regime lies in the smaller value of $U/t$.
Apparently, this region contains the parameter range of $U/t$ and
$\delta$ where perturbation in $U$ is justified.
Therefore, for the states with $\phi = 0$, perturbation theory is
justified.
Although we cannot say whether perturbation theory is reliable in the 
$\phi \neq 0$ regime, states with $\phi \neq 0$ are qualitatively
different from those with $\phi = 0$ because the former is unstable
in the $U/t \rightarrow 0$ limit.
(In the language of renormalization group theory,
they should belong to different fixed points.)
Turning to the conditions of d-wave superconductivity, the boundaries
of $\Delta_c/t=0.20$ and $\Delta_c/t=0.04$ in Fig.~\ref{fig_af}
suggests that the occurrence of $d$-wave superconductivity is
restricted to extremely small parameter region or we need much
stronger attractive interaction than the RKKY type interaction.

{\it Another $d$-wave pairing-}
So far we discuss the possibility of $d$-wave superconductivity in 
the carrier system.
Now we discuss that there is another $d$-wave ($d_{x^2-y^2}$-wave)
pairing associated with the localized spin moments.

In the $\phi \neq 0$ regime, we may take 
the form of the antiferromagnetic Heisenberg Hamiltonian 
for the action of the localized spin moments.
In the Hamiltonian formulation, it is written as 
$H_{\rm spin} = \frac{J\phi^2}{4} \sum_{\langle i,j \rangle}
{\mbox{\boldmath ${\bf \Omega}$}}_i \cdot
{\mbox{\boldmath ${\bf \Omega}$}}_j$, with
$J=4t^2/U$.\cite{SCHULZ,LACOUR-GAYET_CYROT}
Note that the exchange interaction between the localized spins is
reduced by factor $\phi^2$. Since $\phi$ is monotonically
decreasing function of the doped carrier concentration $\delta$,
this reduction suggests the relation to the Heisenberg antiferromagnet
like behavior of spin susceptibility observed in the doped high $T_c$
cuprates. \cite{JOHNSTON}

Now we discuss another $d$-wave pairing.
In order to describe the localized spin moments 
${\mbox{\boldmath ${\bf \Omega}$}}_j$, we can introduce fermion
creation and annihilation operators, $a_j^{\dagger}$ and $a_j$, as 
${\mbox{\boldmath ${\bf \Omega}$}}_j
=a_j^{\dagger} {\mbox{\boldmath ${\bf \sigma}$}} a_j$.
Due to the constraint $|{\mbox{\boldmath ${\bf \Omega}$}}_j|=1$, 
the fermion operators $a_j^{\dagger}$ and $a_j$ must satisfy,
\begin{equation}
\sum_{\sigma} a_{j\sigma}^{\dagger} a_{j\sigma}=1.
\label{eq_constraint}
\end{equation}
Note that this constraint, the fermion system is half-filled, is
independent of the doping concentration $\delta$. 
Under the constraint (\ref{eq_constraint}), the Hamiltonian $H_{\rm
spin}$ is reduced to, up to a constant term,
\begin{equation}
H_{\rm spin} = - \frac{J\phi^2}{2} \sum_{\langle i,j \rangle}
D_{ij}^{\dagger} D_{ij},
\end{equation}
where $D_{ij} = a_{i\uparrow}a_{j\downarrow}
- a_{i\downarrow}a_{j\uparrow}$ is defined on each bond.
Taking the mean fields $\langle D_{ij} \rangle$
and $\langle D_{ij}^{\dagger} \rangle$, 
we find that the $d_{x^2-y^2}$-wave
pairing state and the extended $s$-wave state are degenerate.
If we introduce a slight hopping term for the fermions,
then the $d_{x^2-y^2}$-wave pairing state is
stabilized\cite{KOTLIAR_LIU} and the gap is of order of $\phi^2 J$.
Although the origin of this $d_{x^2-y^2}$ pairing is similar to the
spinon pairing in the RVB theory, that is, the short-range
antiferromagnetic correlation\cite{ANDERSON_ETAL}, the crucial
difference is that we do not rely on the $U(1)$ 
gauge symmetry breaking that is essential for the slave-particle gauge 
theory.\cite{NAYAK}
In addition, it should be stressed that this pairing state does not
imply a superconducting state of the fermions because of the
constraint (\ref{eq_constraint}).

This pairing state provides another $d_{x^2-y^2}$
pairing which is independent of $d_{x^2-y^2}$-wave superconductivity.
The fact that this pairing state originates from the antiferromagnetic 
correlation between the localized spin moments suggests that it can be 
associated with the pseudogap phenomenon observed in the high $T_c$
cuprates.
If we apply the theory to the $Cu$ site degrees of freedom in the
$CuO_2$ plane in the cuprates, then the $d_{x^2-y^2}$
pairing can be identified with that observed by angle-resolved
photoemission spectroscopy.\cite{ARPES}
Furthermore, there is the experiment that indicates the existence of
the pseudogap of $d_{x^2-y^2}$ symmetry also in the insulating
phase.\cite{RONNING_ETAL}
Since our $d_{x^2-y^2}$ pairing exists also in the insulating
phase, the experiment supports the relationship between experimentally
observed pseudogap with $d_{x^2-y^2}$ symmetry and the $d_{x^2-y^2}$
pairing due to the antiferromagnetic correlation.

{\it Summary-}
To summarize, we have reexamined the possibility of $d$-wave
superconductivity in the hole-doped Hubbard model.
We have shown that $d$-wave superconductivity is unstable in the
perturbative region in $U$. 
Whereas in the region where $d$-wave superconductivity is possible, 
there is a strong constraint on the superconducting gap.
Instead, there is another $d_{x^2-y^2}$-wave pairing due to the
antiferromagnetic spin correlation.

{\it Acknowledgement-}
I would like to thank M. Sigrist, Y. Morita, and M. Tsuchiizu for
helpful discussions.
This work was supported in part by a Grant-in-Aid from the Ministry of 
Education, Culture, Sports, Science and Technology.

%


\end{multicols}

\begin{references}
\bibitem{SPIRAL} As shown in Ref.~\cite{SCHULZ}, spiral phase is
expected away from half-filling because the kinetic energy of carriers 
enhances ferromagnetic interaction between the localized spins. In
order to concentrate on d-wave superconductivity based on
antiferromagnetic correlations, the staggered component is taken here.
\end{references}
\end{document}